\documentclass[aps,prd,onecolumn,superscriptaddress,altaffilletter,nobibnotes,nofootinbib,10pt,showpacs,showkeys,preprintnumbers]{revtex4-2}

\usepackage{float}
\usepackage{graphicx}
\usepackage{dcolumn}
\usepackage[dvipsnames]{xcolor}
\usepackage{bm}
\usepackage{lineno}
\usepackage{float}
\usepackage{amsmath}
\usepackage{soul}
\usepackage{tikz,xcolor,hyperref}

\definecolor{lime}{HTML}{A6CE39}
\DeclareRobustCommand{\orcidicon}{
	\begin{tikzpicture}
	\draw[lime, fill=lime] (0,0) 
	circle [radius=0.16] 
	node[white] {{\fontfamily{qag}\selectfont \tiny ID}};
	\draw[white, fill=white] (-0.0625,0.095) 
	circle [radius=0.007];
	\end{tikzpicture}
	\hspace{-2mm}
}
\foreach \x in {A, ..., Z}{\expandafter\xdef\csname orcid\x\endcsname{\noexpand\href{https://orcid.org/\csname orcidauthor\x\endcsname}
			{\noexpand\orcidicon}}
}

\begin{document}


\title{A unified formalism to study $soft$ as well as $hard$ part of the transverse momentum spectra}
\author{Rohit Gupta}
\affiliation{Department of Physical Sciences, Indian Institute of Science Education and Research (IISER) Mohali, Sector 81 SAS Nagar, Manauli PO 140306 Punjab, India}


\author{Satyajit Jena}
\email{sjena@iisermohali.ac.in}
\affiliation{Department of Physical Sciences, Indian Institute of Science Education and Research (IISER) Mohali, Sector 81 SAS Nagar, Manauli PO 140306 Punjab, India}

\begin{abstract}
Transverse momentum $p_T$ spectra of final state particles produced in high energy heavy-ion collision can be divided into two distinct regions based on the difference in the underlying particle production process. We have provided a unified formalism to explain both low- and high-$p_T$ regime of spectra in a consistent manner. The $p_T$ spectra of final state particles produced at RHIC and LHC energies have been analysed using unified formalism to test its applicability at different energies, and a good agreement with the data is obtained across all energies. Further, the prospect of extracting the elliptic flow coefficient directly from the transverse momentum spectra is explored.
\end{abstract}

\maketitle

\section{Introduction}
Among the primary goals of heavy-ion collision experiments are to understand the QCD matter produced under extreme  temperature and energy density called Quark-Gluon Plasma (QGP).  Experiments at RHIC and LHC are trying to explore the QCD phase diagram in a quest to search for the confinement-deconfinement phase transition and the QCD critical point. \\
Constraints on the detector capabilities and the short time scale of the collision limit the direct search for this new state of strongly interacting quarks and gluons. Hence, we rely on the information carried by the final state particles to study the initial stages of high energy collision. Kinematics observables of the final state particles such as the transverse momenta $p_T$ and pseudorapidity $\eta$ play crucial role in understanding the dynamics of particle production. Quantum Chromodynamics (QCD) is the theoretical framework to study the strongly interacting particles; however the asymptotic freedom in the coupling strength of the theory limits the applicability of the perturbative theory only at higher energy as it breaks at lower energy. Hence, we utilise the statistical thermal models to study the spectra of particles produced in such collision.  \\
In the present work, we will discuss a unified formalism  developed to study the transverse momentum spectra in the heavy-ion collision. \\
In section (2) we will discuss the theoretical framework that are being used to study the transverse momentum spectra. The unified formalism that we have developed to study the spectra is discussed in section (3) followed by the result and conclusion in section (4) and (5) respectively.
\section{Conventional Approach}
Considering a purely thermal source, we can use standard Boltzmann-Gibbs (BG) statistics to explain the energy distribution of the particles. Formula governing the energy distribution in case of BG statistics will be of the form of negative exponential of energy. The distribution function for transverse momentum in BG statistics is given as:
\begin{equation}
\frac{1}{2\pi p_T} \frac{d^2 N}{dp_T dy} = \frac{gV m_T}{(2 \pi)^3}  exp\left(- \frac{m_T}{T}\right)  
\label{eq:bg}
\end{equation}
It is observed that the BG distribution function deviates heavily from the data indicating that the modification is required in standard BG formalism to explain the spectra.
Application of BG statistics is limited to the extensive systems where entropy is additive; however, there exist some strongly correlated non-extensive systems where entropy can be non-additive.\\
Tsallis statistics \cite{Tsallis:1987eu} is proposed as a generalization to standard BG statistics with an additional parameter taking care of non-extensivity in the system. Tsallis statistics modifies normal exponential into a $q$-exponential where $'q'$ is the non-extensitivity parameter. The distribution function for transverse momenta in Tsallis formalism is of the form:
\begin{equation}
\frac{1}{2\pi p_T} \frac{d^2 N}{dp_T dy} = \frac{gV m_T}{(2 \pi)^3} \left[1+(q-1)\frac{m_T}{T}\right]^{-\frac{q}{q-1}} 
\label{eq:tsa}
\end{equation}
Tsallis distribution function given in Eq. (\ref{eq:tsa}) has been used extensively to explain the spectra in low $p_T$ region.\\
Transverse momentum spectra of particles produced in the heavy-ion collision can be divided into two distinct regions depending on the difference in the underlying process of particle production. The low-$p_T$ region consists of particle produced in soft processes whereas the particles produced in hard-scattering processes populates the high-$p_T$ part of the spectra. The low-$p_T$ part of the spectra is explained using the Tsallis statistics, whereas pQCD based power-law form of the function is used for the high-$p_T$ region. \\
We have developed a unified formalism to explain both soft and hard processes in a consistent manner.
\section{Unified Statistical Framework}
A generalized distribution function defined in terms of the first four moments related to mean, sigma, skewness and kurtosis of the distribution has been provided by Karl Pearson \cite{Pearson343} in 1895. It is described in terms of a differential equation:
\begin{equation}\label{diffe}
 \frac{1}{p(x)}\frac{dp(x)}{dx} + \frac{a + x}{b_0 + b_1 x + b_2 x^2} = 0
\end{equation}
and under different limit on its parameters $a, b_0, b_1, b_2$, it reduces to Gaussian, Cauchy, gamma, inverse-gamma and other distribution \cite{pollard}.
Solution to differential equation Eq.~(\ref{diffe}) is given as:
\begin{equation}\label{PEARSON}
p(x) = B \left( 1+ \frac{x}{e}\right)^f  \left( 1 + \frac{x}{g}\right)^h   
\end{equation}
This formalism is used to develop a unified framework Ref.~\cite{Jena:2020wno} and the function for transverse momentum in this model is given as:
\begin{equation}\label{pearson_final}
    \frac{1}{2\pi p_T} \frac{d^2 N}{dp_T dy} = B'  \left( 1 + \frac{p_T}{p_0}\right)^{-n} \left( 1 + (q-1)\frac{p_T}{T}\right)^{-\frac{q}{q-1}} 
\end{equation}
Backward compatibility to Tsallis statistics and thermodynamical consistency of this unified formalism has been proved in Ref.~\cite{Gupta:2020naz}. 
\section{Result and conclusion}
 \begin{figure}[h!]
 \centering
  \begin{minipage}[b]{0.4\textwidth}
    \includegraphics[width=\textwidth]{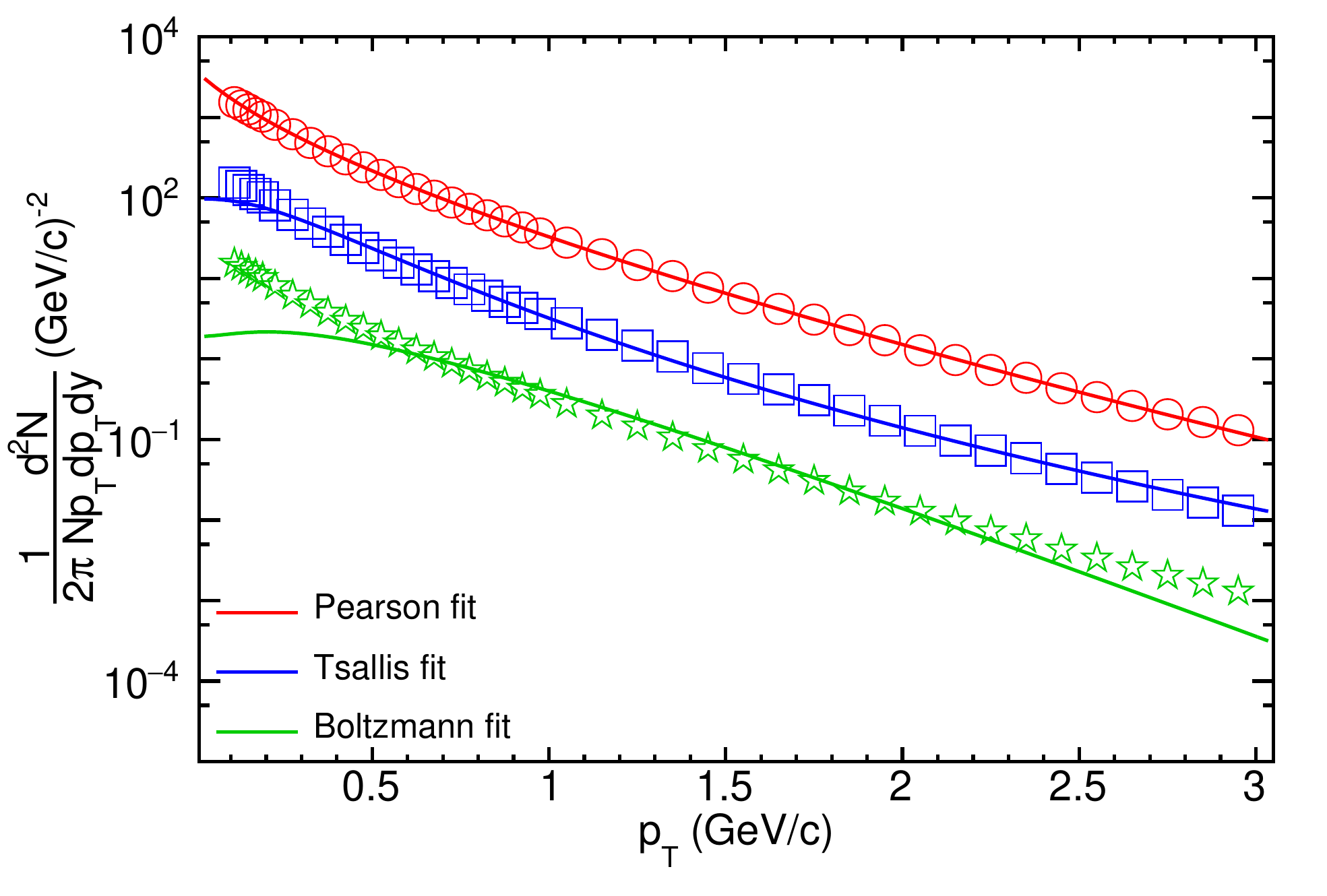}
    \caption{The $p_T$ spectra of $\pi^+$ produced in most central 2.76 TeV $PbPb$ collision \cite{Adler:2003cb} fitted with Boltzmann, Tsallis and Pearson distribution function. Data points are scaled for better visibility.}
    \label{fig:2760}
  \end{minipage}
  \hfill
  \begin{minipage}[b]{0.4\textwidth}
    \includegraphics[width=\textwidth]{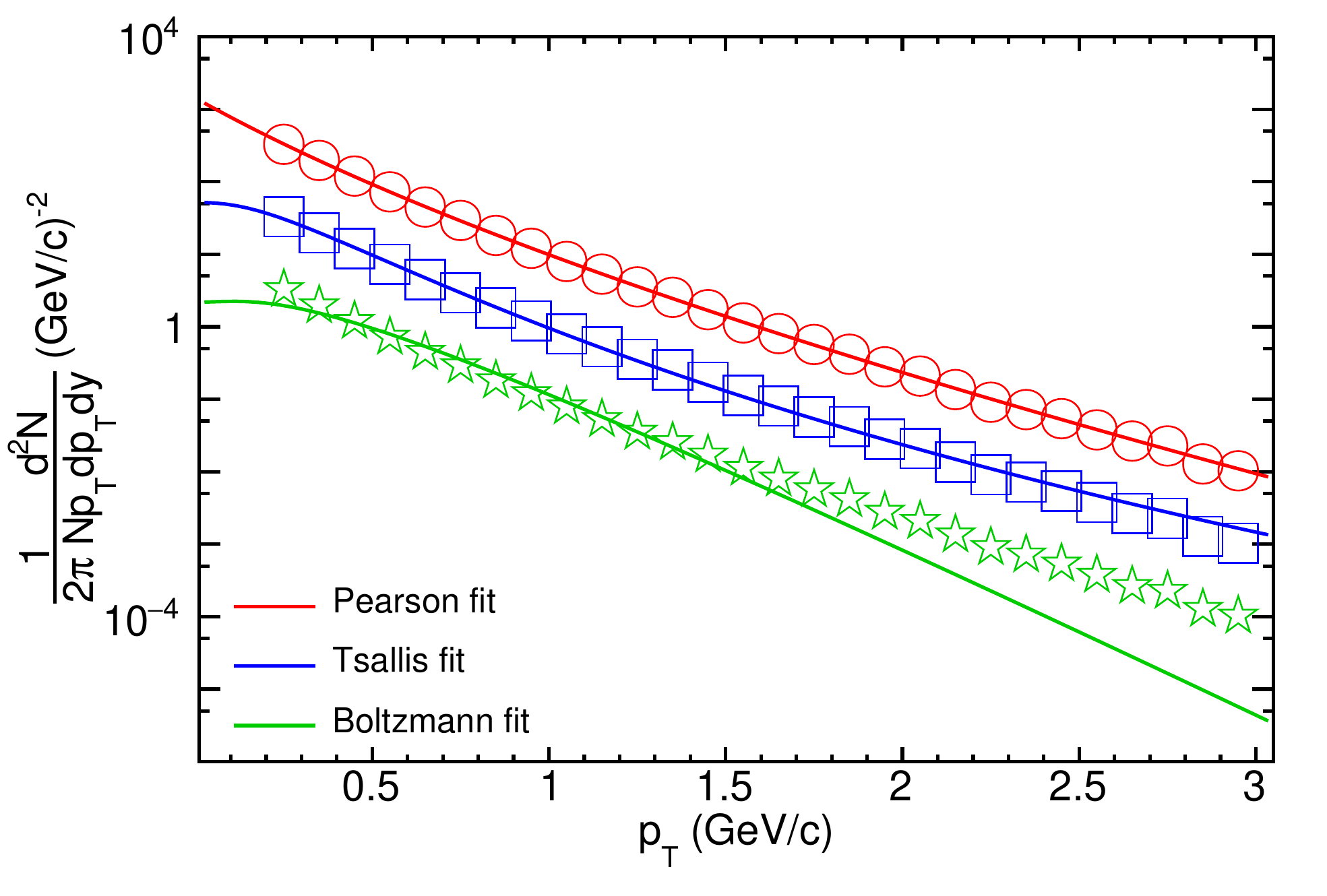}
    \caption{The $p_T$ spectra of $\pi^+$ produced in most central 200 GeV $AuAu$ collision \cite{Abelev:2013vea} fitted with Boltzmann, Tsallis and Pearson distribution function. Data points are scaled for better visibility.}
    \label{fig:200}
  \end{minipage}
  \end{figure}
\begin{figure}[h!]
\centering
  \includegraphics[height=2in,width=3in]{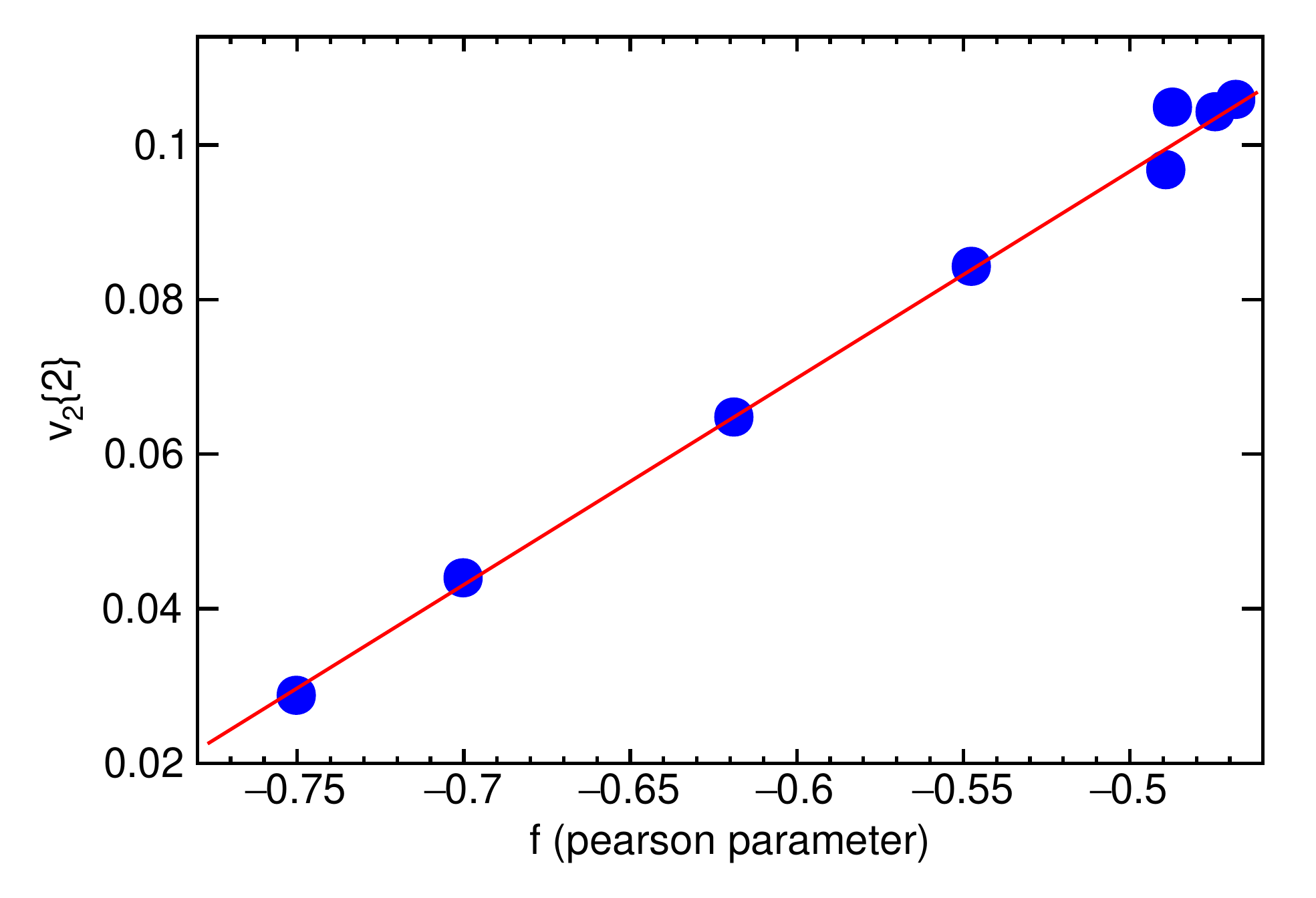}
  \caption{ Flow coefficient $v_2$ versus Pearson parameter f at $2.76$ TeV $PbPb$ collision. Each data point belong to one centrality class.}\label{fig:v2vsf}
\end{figure}
We have tested the applicability of unified formalism by analysing transverse momentum spectra of particles produced in heavy-ion collision at different energies. In Fig.~(\ref{fig:2760}, \ref{fig:200}), we have plotted the $p_T$ spectra of positive pions produced in $2.76$ TeV $PbPb$ and $200$ GeV $AuAu$ collision. Solid lines correspond to the final fit to different distribution functions obtained using the $\chi^2$-minimization technique. The goodness of fit is measured in terms of $\chi^2/NDF$ and the values obtained for the unified formalism show a good agreement between the data and the fit function. \\
We extended our analysis to search to the connection of two parameters $p_0$ and $n$ with the observables in collider experiments. A similarity in the centrality dependence of the second-order flow coefficient $v_2$ \cite{Aamodt:2010pa} and the Pearson parameter $f (=-n)$ is observed, indicating a connection between two quantities. We obtained a linear relation between the two parameters for charged hadrons data at $2.76$ TeV as shown in Fig.~(\ref{fig:v2vsf}). \\
In conclusion, a unified formalism has been discussed to study the transverse momentum spectra at different energies and the relation between the Pearson parameter and the flow coefficient is established.

%

\end{document}